\documentclass[a4paper,11pt]{article}
\pdfoutput=1

\def\be{\begin{equation}}
\def\ee{\end{equation}}
\def\bea{\begin{eqnarray}}
\def\eea{\end{eqnarray}}

\usepackage{
graphicx,
hyperref,
amsmath,
amssymb,
charter,
xcolor,
ifluatex,
longtable,
booktabs,
multirow,
bbold,cancel}

\usepackage{
graphicx,
hyperref,
amsmath,
amssymb,
charter,
xcolor,
ifluatex,
longtable,
booktabs,
multirow,
bbold}
\usepackage{colortbl}

\definecolor{Gray}{gray}{0.95}
\definecolor{RGray}{gray}{0.85}
\definecolor{CGray}{gray}{0.92}

\usepackage{multicol}
\definecolor{tit}{rgb}{0.1,0.2,0.4}
\definecolor{blus}{cmyk}{1,1,0,0.6}
\definecolor{verde}{cmyk}{0.92,0,0.59,0.25}

\usepackage{tikz}
\usetikzlibrary{matrix}

\usepackage{devanagari}
\usepackage{tipa}
\usepackage{amsmath,amssymb,amsfonts, bm}
\usepackage{dsfont}
\usepackage{epsfig}
\usepackage{graphicx}
\usepackage{slashed}
\usepackage{color}

\usepackage{caption}
\usepackage{subcaption}
\usepackage{slashed} 

\usepackage{adjustbox}
\usepackage{commath}
\usepackage{calc}

\usepackage{sidecap}

\newcommand{\bfig}{\begin{figure}}
\newcommand{\efig}{\end{figure}}

\usepackage{slashed}

\usepackage{commath}
\usepackage{calc}

\newcommand{\D}{{\cal D}}
\newcommand{\U}{{\cal U}}

\newcommand{\M}{{\cal M}}

\makeatletter
\newcommand*{\rom}[1]{\expandafter\@slowromancap\romannumeral #1@}
\makeatother

\usepackage{color,hyperref}
\definecolor{darkblue}{rgb}{0.0,0.0,0.3}
\hypersetup{colorlinks,breaklinks,
            linkcolor=darkblue,urlcolor=darkblue,
            anchorcolor=darkblue,citecolor=darkblue}

\def\1by3{\ensuremath{\frac{1}{3}}}
\def\4by3{\ensuremath{\frac{4}{3}}}
\def\2by3{\ensuremath{\frac{2}{3}}}

\providecommand*\url[1]{\href{#1}{#1}}
\renewcommand*\url[1]{\href{#1}{\texttt{#1}}}

\begin{document}

\allowdisplaybreaks
\vspace*{-2.5cm}
\begin{flushright}
{\small
IIT-BHU
}
\end{flushright}

\vspace{2cm}

\begin{center}
{\LARGE \bf \color{tit}
 Discrete origins of matter}\\[1cm]

{\large\bf Gauhar Abbas$^{a}$\footnote{email: gauhar.phy@iitbhu.ac.in}   }  
\\[7mm]

{\large\bf Neelam Singh$^{a}$\footnote{email: neelamsingh.rs.phy19@itbhu.ac.in}   }  \\[7mm]

{\it $^a$ } {\em Department of Physics, Indian Institute of Technology (BHU), Varanasi 221005, India}\\[3mm]

\vspace{1cm}

{\large\bf\color{blus} Abstract}
\begin{quote}
We discuss models of the flavour problem and dark matter based on the discrete  $\mathcal{Z}_{\rm N} \times \mathcal{Z}_{\rm M} \times \mathcal{Z}_{\rm P}$ flavour symmetry.  A new class of dark-matter emerges out of these models, which is defined as the flavonic dark matter.  An ultra-violet completion of these models based on the dark-technicolour paradigm is also presented.
\end{quote}

\thispagestyle{empty}
\end{center}

\begin{quote}
{\large\noindent\color{blus} 
}

\end{quote}

\newpage
\setcounter{footnote}{0}

\section{Introduction}
\label{intro}
The incompleteness of the standard model (SM) manifests itself in many ways.  For instance, the absence of any mechanism for the origin of  the fermionic mass spectrum and the mixing patterns, including that of leptons.  This is known as the flavour problem.  For a review of the flavour problem, see reference \cite{Abbas:2023ivi}.  Moreover, discovery of dark matter extends the span of the incompleteness of the SM to a point where the need to go beyond the SM becomes inevitable.

In this work, we discuss some new developments to solve the problem of flavour and dark matter together through a generic   $\mathcal{Z}_{\rm N} \times \mathcal{Z}_{\rm M} \times \mathcal{Z}_{\rm P}$ discrete flavour  symmetry  \cite{Abbas:2017vws,Abbas:2018lga,Abbas:2020frs}.  The ultra-violet origin of this symmetry lies in a dark-technicolour scenario where multi-fermion chiral condensates provide a solution of the flavour problem  \cite{Abbas:2017vws,Abbas:2020frs,Abbas:2023dpf,Abbas:2023bmm}.  A subset of this symmetry,  $\mathcal{Z}_{\rm N} \times \mathcal{Z}_{\rm M}$,  is capable of providing a dark matter candidate referred to as the `` flavonic dark matter" together with a solution to the flavour problem \cite{Abbas:2023ion}.

We shall present our discussion along the following track: In section \ref{section2}, we present a phenomenological investigation of two proto-type $\mathcal{Z}_{\rm N} \times \mathcal{Z}_{\rm M}$ flavour symmetries by deriving the bounds of the flavon field of the   Froggatt-Nielsen (FN) mechanism \cite{Froggatt:1978nt} using the flavour physics data.  In section \ref{sec3}, we discuss a joint solution of the flavour problem and dark matter, which gives rise to the flavonic dark matter.  We discuss an atypical solution to the flavour problem based on the VEV hierarchy in section \ref{sec4}, which is based on the  $\mathcal{Z}_{\rm N} \times \mathcal{Z}_{\rm M} \times \mathcal{Z}_{\rm P}$ flavour symmetry. A dark-technicolour paradigm providing a UV completion of the  $\mathcal{Z}_{\rm N} \times \mathcal{Z}_{\rm M} \times \mathcal{Z}_{\rm P}$ flavour symmetry  and the VEV hierarchy is discussed in section \ref{sec5}.

\section{{ $\mathcal{Z}_{\rm N} \times \mathcal{Z}_{\rm M}$ flavour symmetry}}
\label{section2}
The $\mathcal{Z}_{\rm N} \times \mathcal{Z}_{\rm M}$ flavour symmetry  effectively determines flavour structure of the SM, including that of neutrinos \cite{Abbas:2018lga}.  The key idea is to extend SM symmetry by introducing a complex singlet  scalar flavon field  $\chi$,  which transforms under the $SU(3)_c \times SU(2)_L \times U(1)_Y$ symmetry of the SM as,
\begin{eqnarray}
\chi :(1,1,0).
 \end{eqnarray} 
The Lagrangian that provides masses to the charged fermions of the SM reads as,
\bea
\label{mass1}
-{\mathcal{L}}_{\rm Yukawa} &=&    \left[  \dfrac{ \chi(\chi^{\dagger})}{\Lambda} \right]^{n_{ij}^u}     y_{ij}^u \bar{ \psi}_{L_i}^q  \tilde{\varphi} \psi_{R_j}^{u}  +   \left[  \dfrac{ \chi(\chi^{\dagger})}{\Lambda} \right]^{n_{ij}^d}      y_{ij}^d \bar{ \psi}_{L_i}^q  \varphi \psi_{R_j}^{d}  \nonumber \\
&+&    \left[  \dfrac{ \chi(\chi^{\dagger})}{\Lambda} \right]^{n_{ij}^\ell}        y_{ij}^\ell \bar{ \psi}_{L_i}^\ell  \varphi \psi_{R_j}^{\ell} 
+  {\rm H.c.}, \\ \nonumber
&=&  Y^u_{ij} \bar{ \psi}_{L_i}^q  \tilde{\varphi} \psi_{R_j}^{u}
+ Y^d_{ij} \bar{ \psi}_{L_i}^q  \varphi \psi_{R_j}^{d}
+ Y^\ell_{ij} \bar{ \psi}_{L_i}^\ell  \varphi \psi_{R_j}^{\ell}   + \text{H.c.}. 
\eea

Employing the principle of minimum suppression \cite{Abbas:2022zfb}, it turns out that $\mathcal{Z}_{2} \times \mathcal{Z}_{5}$ is the simplest realization of  $\mathcal{Z}_{\rm N} \times \mathcal{Z}_{\rm M}$ symmetry. We term it as the minimal $\mathcal{Z}_{\rm N} \times \mathcal{Z}_{\rm M}$  model. Under this symmetry, the charges to various fields are assigned in the way, as depicted in Table \ref{tab_charges}. 

\begin{table}[h]
\begin{center}
\begin{tabular}{|c|c|c|c|c|c||c|c|c|c|c|c|}
  \hline
  Fields             &        $\mathcal{Z}_2$                    & $\mathcal{Z}_5$ & Fields             &        $\mathcal{Z}_2$                    & $\mathcal{Z}_5$ & Fields             &        $\mathcal{Z}_2$                    & $\mathcal{Z}_9$ & Fields             &        $\mathcal{Z}_2$                    & $\mathcal{Z}_9$        \\
  \hline
  $u_{R}, c_{R}, t_{R}$                 &   +  & $ \omega^2$      & $\psi_{L_2}^q$,  $\psi_{L_2}^\ell$                   &   +  &     $\omega^4 $       &  $u_{R}, t_{R}$                 &   +  & $ 1$       & $\psi_{L_1}^q$,  $\psi_{L_1}^\ell$                   &   +  &    $\omega $             \\
   $d_{R},  s_{R}, b_{R}$                 &   -  &     $\omega $           &  $\psi_{L_3}^q$,  $\psi_{L_3}^\ell$                &   +  &      $ \omega^2 $      &  $c_{R}$                 &   +  & $ \omega^4$   &  $\psi_{L_2}^q$,  $\psi_{L_2}^\ell$                  &   +  &     $\omega^8 $            \\
  $ e_R, \mu_R, \tau_R$                 &   -  &     $\omega $              &  $\chi$                        & -  &       $ \omega$      & $d_{R},  s_{R},  b_{R}$                &   -  &     $\omega^3 $    &  $\psi_{L_3}^q$,    $\varphi$                   &   +  &      $ 1 $              \\
  $ \nu_{e_R},   \nu_{\mu_R}, \nu_{\tau_R} $                 &   -  &     $\omega^3 $                              &  $\varphi$              &   +        &     1 & $ e_R, \mu_R, \tau_R $                 &   -  &     $\omega^3 $ & $\psi_{L_3}^\ell$                 &   +  &      $ \omega^6 $     \\
   $\psi_{L_1}^q$,    $\psi_{L_1}^\ell$                   &   +  &    $\omega $                          &              &           &  &  $ \nu_{e_R},   \nu_{\mu_R},  \nu_{\tau_R}  $                 &   -  &     $\omega^7 $ & $\chi$                        & -  &       $ \omega$     \\
  \hline
     \end{tabular}
\end{center}
\caption{Assigned charges to SM and flavon fields under $\mathcal{Z}_2$, $\mathcal{Z}_5$, and $\mathcal{Z}_9$  symmetries,  where $\omega$ is the fifth or ninth root of unity. }
\label{tab_charges}
\end{table}
It is observed that certain Yukawa couplings in the minimal model based on $\mathcal{Z}_{2} \times \mathcal{Z}_{5}$ flavour symmetry are not order-one, which is conventionally favoured in literature. We adopt a non-minimal form of $\mathcal{Z}_{\rm N} \times \mathcal{Z}_{\rm M}$ paradigm, $\mathcal{Z}_{2} \times \mathcal{Z}_{9}$ flavour symmetry, where the Yukawa couplings are indeed order-one, and assign charges to the SM and flavon fields, as shown in Table \ref{tab_charges}.

\subsection{Masses and mixing patterns}
The minimal model based on $\mathcal{Z}_{2} \times \mathcal{Z}_{5}$ flavour symmetry allows following mass matrices for up and down-type quarks, and charged leptons,
\begin{equation}
\M_u = \dfrac{v}{\sqrt{2}}
\begin{pmatrix}
y_{11}^u  \epsilon^4 &  y_{12}^u \epsilon^4  & y_{13}^u \epsilon^4    \\
y_{21}^u  \epsilon^2    & y_{22}^u \epsilon^2  &  y_{23}^u \epsilon^2    \\
y_{31}^u     &  y_{32}^u      &  y_{33}^u
\end{pmatrix}, 
\M_d = \dfrac{v}{\sqrt{2}}
\begin{pmatrix}
y_{11}^d  \epsilon^5 &  y_{12}^d \epsilon^5 & y_{13}^d \epsilon^5   \\
y_{21}^d  \epsilon^3  & y_{22}^d \epsilon^3 &  y_{23}^d \epsilon^3  \\
 y_{31}^d \epsilon &  y_{32}^d \epsilon   &  y_{33}^d \epsilon
\end{pmatrix}, 
\M_\ell =  \dfrac{v}{\sqrt{2}}
\begin{pmatrix}
y_{11}^\ell  \epsilon^5 &  y_{12}^\ell \epsilon^5  & y_{13}^\ell \epsilon^5   \\
y_{21}^\ell  \epsilon^3  & y_{22}^\ell \epsilon^3  &  y_{23}^\ell \epsilon^3  \\
 y_{31}^\ell \epsilon   &  y_{32}^\ell \epsilon   &  y_{33}^\ell \epsilon
\end{pmatrix}.
\end{equation}
And for the non-minimal $\mathcal{Z}_{2} \times \mathcal{Z}_{9}$ flavour symmetry, the corresponding mass matrices are 
\begin{equation}
M_u = \dfrac{v}{\sqrt{2}}
\begin{pmatrix}
y_{11}^u  \epsilon^8 &  y_{12}^u \epsilon^{6}  & y_{13}^u \epsilon^{8}    \\
y_{21}^u  \epsilon^8    & y_{22}^u \epsilon^4  &  y_{23}^u \epsilon^8   \\
y_{31}^u      &  y_{32}^u  \epsilon^4     &  y_{33}^u  
\end{pmatrix},
M_d = \dfrac{v}{\sqrt{2}}
\begin{pmatrix}
y_{11}^d  \epsilon^7 &  y_{12}^d \epsilon^7 & y_{13}^d \epsilon^7   \\
y_{21}^d  \epsilon^5  & y_{22}^d \epsilon^5 &  y_{23}^d \epsilon^5  \\
 y_{31}^d \epsilon^3 &  y_{32}^d \epsilon^3   &  y_{33}^d \epsilon^3
\end{pmatrix},
M_\ell =  \dfrac{v}{\sqrt{2}}
\begin{pmatrix}
y_{11}^\ell  \epsilon^7 &  y_{12}^\ell \epsilon^7  & y_{13}^\ell \epsilon^7   \\
y_{21}^\ell  \epsilon^5  & y_{22}^\ell \epsilon^5  &  y_{23}^\ell \epsilon^5  \\
 y_{31}^\ell \epsilon^3   &  y_{32}^\ell \epsilon^3   &  y_{33}^\ell \epsilon^3 
\end{pmatrix}.
\end{equation} 
The masses of the quarks and charged leptons are obtained in terms of the expansion parameter $\epsilon$. Their expressions at the leading-order take up the form as shown in table \ref{tab_masses} \cite{Rasin:1998je} \cite{Abbas:2022zfb}.

\begin{table}[h] 
\begin{center}
\begin{tabular}{|c||c|c|}
\hline 
Masses & $\mathcal{Z}_2 \times \mathcal{Z}_5$  &  $\mathcal{Z}_2 \times \mathcal{Z}_9$  \\ \hline \hline
$\{m_t, m_c, m_u\}$ & $\simeq \{|y_{33}^u| , ~  |y_{22}^u| \epsilon^2, ~  |y_{11}^u| \epsilon^4\}v/\sqrt{2}$  & $\simeq \{|y_{33}^u| , ~  |y_{22}^u| \epsilon^4, ~  |y_{11}^u| \epsilon^8\}v/\sqrt{2} $ \\ 
$\{m_b, m_s, m_d\}$ & $\simeq \{|y_{33}^d| \epsilon, ~ |y_{22}^d| \epsilon^3, ~  |y_{11}^d| \epsilon^5\}v/\sqrt{2}$  & $\simeq \{|y_{33}^d| \epsilon^3, ~ |y_{22}^d| \epsilon^5 ,~  |y_{11}^d| \epsilon^7\}v/\sqrt{2}$ \\ 
$\{m_\tau, m_\mu, m_e\}$  & $\simeq \{|y_{33}^l| \epsilon, ~ |y_{22}^l| \epsilon^3, ~   |y_{11}^l| \epsilon^5\}v/\sqrt{2}$  & $\simeq \{|y_{33}^l| \epsilon^3, ~ |y_{22}^l| \epsilon^5, ~   |y_{11}^l| \epsilon^7\}v/\sqrt{2}$ \\ \hline
\end{tabular}
\caption{Leading order mass patterns of quarks and leptons in terms of parameter $\epsilon$ for minimal and non-minimal models.}
\label{tab_masses}  
\end{center}
\end{table} 

The quark mixing angles up to the leading order are  approximately as given in table \ref{tab_mixing1} \cite{Rasin:1998je}.
\begin{table}[h]
\begin{center}
\begin{tabular}{|c||c|c|}
\hline 
Quark mixing angles & $\mathcal{Z}_2 \times \mathcal{Z}_5$  &  $\mathcal{Z}_2 \times \mathcal{Z}_9$  \\ \hline \hline
$\sin \theta_{12}  \simeq |V_{us}|$ & $\simeq \left|{y_{12}^d \over y_{22}^d}  -{y_{12}^u \over y_{22}^u}  \right| \epsilon^2 $ & $\simeq \left|{y_{12}^d \over y_{22}^d}  -{y_{12}^u \over y_{22}^u}  \right| \epsilon^2 $\\ 
 $\sin \theta_{23}  \simeq |V_{cb}|$ & $\simeq |{y_{23}^d \over y_{33}^d}  -{y_{23}^u \over y_{33}^u}  |  \epsilon^2 $ & $\simeq \left|{y_{23}^d \over y_{33}^d}  \right|  \epsilon^2 $ \\ 
$\sin \theta_{13}  \simeq |V_{ub}|$  & $ \simeq \left|{y_{13}^d \over y_{33}^d}  -{y_{12}^u y_{23}^d \over y_{22}^u y_{33}^d} - {y_{13}^u \over y_{33}^u} \right|  \epsilon^4$  & $\simeq \left|{y_{13}^d \over y_{33}^d}  -{y_{12}^u y_{23}^d \over y_{22}^u y_{33}^d} \right|  \epsilon^4 $ \\ \hline
\end{tabular}
\caption{Leading order expressions for quark mixing angles in terms of the parameter $\epsilon$ for minimal and non-minimal models.}
\label{tab_mixing1}
\end{center} 
\end{table} 

For our numerical analysis, the values of the expansion parameter $\epsilon$ turn out to be 0.1 and 0.23 for the minimal and non-minimal models, respectively \cite{Abbas:2022zfb}.
\subsection{The scalar potential} 
The scalar potential of the model can be written as,
\begin{align}
- V &= - \mu^2 \varphi^\dagger \varphi +\lambda (\varphi^\dagger \varphi)^2 - \mu_\chi^2\, \chi^*  \chi  + \lambda_\chi\, (\chi^* \chi)^2 + (\rho  \ \chi^2 + \mathrm{H.c.}) + \lambda_{\varphi  \chi} (\chi^* \chi) (\varphi^\dagger \varphi).
\label{eq:potential}
\end{align}
Under the assumption that there is no Higgs-flavon mixing, $ \lambda_{\varphi  \chi}   =0$, flavon field can be parametrized by excitations around its VEV in the following way,
\begin{align}
 \chi(x)=\frac{f + s(x) +i\, a(x)}{\sqrt{2}}.
\label{chi}
\end{align}
where $f$ represents the VEV of the flavon field, while $s$ and $a$ denote its scalar and pseudoscalar components, respectively.

 The minimization conditions for the scalar potential yield the following expressions for the masses of the scalar and pseudoscalar components of the flavon,
\begin{align}
m_s &= \sqrt{\mu_\chi - 2 \rho} = \sqrt{\lambda_\chi} f , \nonumber \\
m_a &= \sqrt{-2 \rho}.
\label{eq:masses}
\end{align}
It is evident from equation \ref{eq:masses} that the mass of the pseudoscalar is contingent on the soft-symmetry breaking parameter $\rho$, establishing it as a free parameter of the model, alongside the VEV $f$.
\subsection{ Bounds on the flavour scale of the $\mathcal{Z}_{\rm 2} \times \mathcal{Z}_{\rm M}$ flavour symmetry }
\begin{figure}[h]
	\centering
	\begin{subfigure}[]{0.36\linewidth}
	 \includegraphics[width=\linewidth]{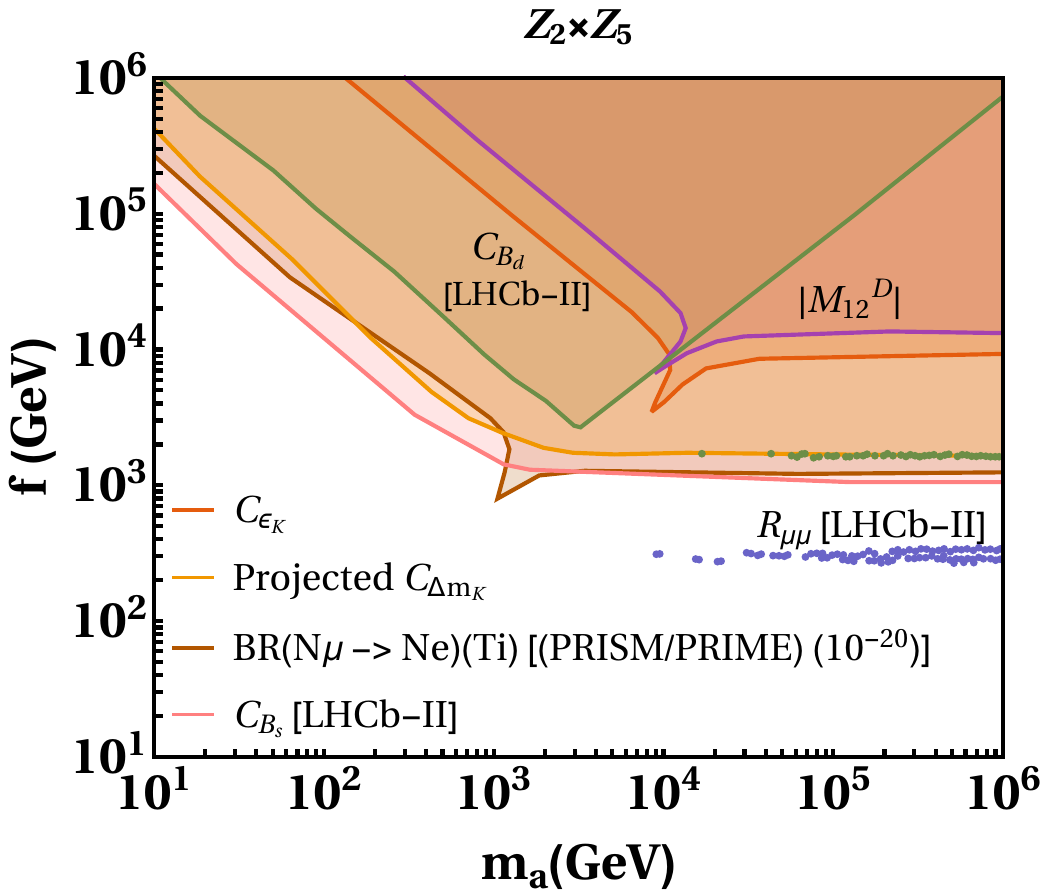}	
	   \caption{}
\end{subfigure} 
\begin{subfigure}[]{0.36\linewidth}
 \includegraphics[width=\linewidth]{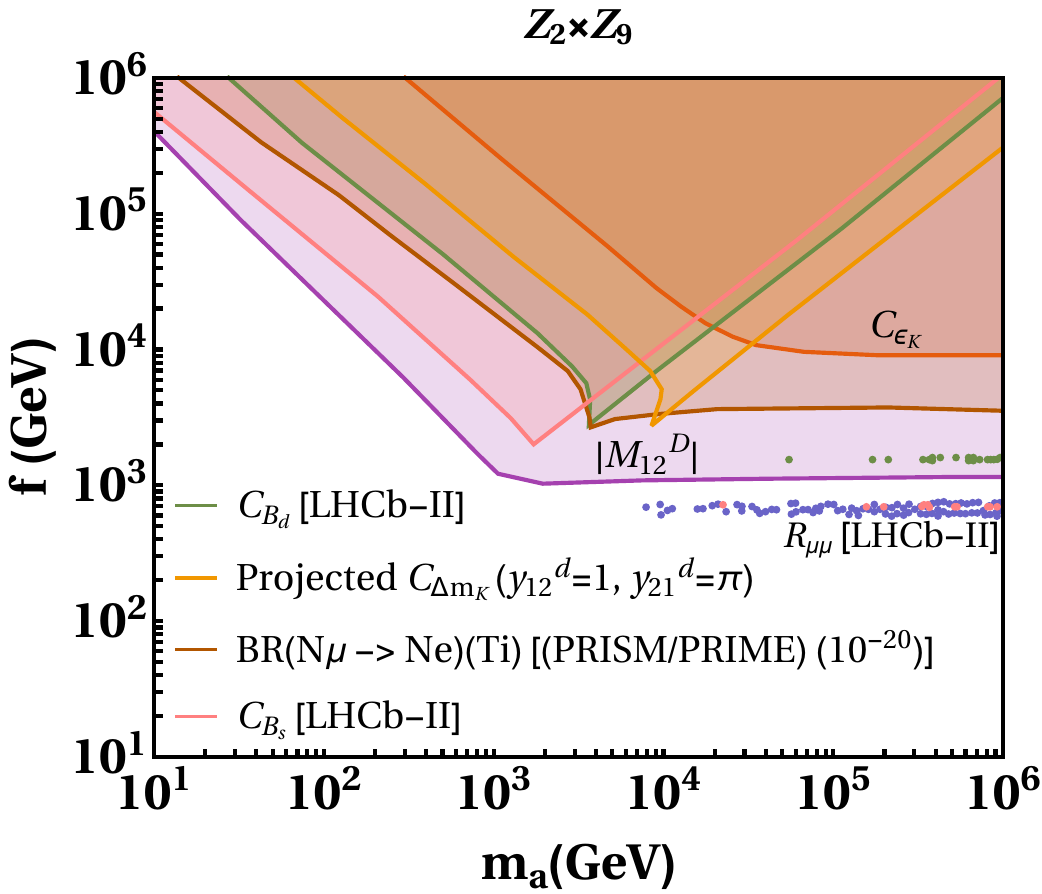}
 \caption{}
 \end{subfigure}
\caption{Summarized representation of constrained $m_a -f$  parameter space for both the minimal $\mathcal{Z}_2 \times \mathcal{Z}_5 $ and non-minimal $\mathcal{Z}_2 \times \mathcal{Z}_9 $ models, depicted in the left and right panels, respectively. }
\label{sum_plot}
\end{figure}

Now we discuss the flavour bounds on the parameter space of the $\mathcal{Z}_{\rm 2} \times \mathcal{Z}_{\rm M}$ flavour symmetries \cite{Abbas:2022zfb}. We show the summary of the most stringent bounds  on the parameter space of both the minimal and non-minimal $\mathcal{Z}_{\rm 2} \times \mathcal{Z}_{\rm M}$ models, where $M= 5,9$, in  figures \ref{sum_plot}.
Currently, the quark flavour observables from $K^0-\bar K^0$ mixing and $D^0 - \bar D^0$ mixing impose stringent constraints on the parameter space for both the minimal and non-minimal models. The observable $R_{\mu \mu}$, which is ratio of $BR(B_{d}\rightarrow \mu^+\mu^-) $ and $BR(B_{s}\rightarrow \mu^+\mu^-) $ turns out to be crucial in the phase-\rom{2} of the LHCb as it eliminates a large parameter space, as shown in figure \ref{sum_plot}.

\section{{ Flavonic dark matter }}
\label{sec3}
The $\mathcal{Z}_{\rm N} \times \mathcal{Z}_{\rm M}$ flavour symmetry framework unveils a remarkable scenario by offering a novel configuration to address both the flavour problem and the existence of dark matter within a unified and comprehensive framework. This is achieved by adding the following term to the scalar  potential \cite{Abbas:2023ion},
\begin{equation} \label{VN}
 V = -\lambda {\chi^{\tilde N} \over \Lambda^{\tilde N-4}} + \text{H.c.}, 
\end{equation}
where  $\lambda=|\lambda|e^{i\alpha}$, and $\tilde N$ denotes the least common multiple of $\rm N$ and $\rm M$ in the $\mathcal{Z}_{\rm N} \times \mathcal{Z}_{\rm M}$ framework. For a value of $\tilde N$ being larger than a particular threshold, the axial flavon field becomes light enough to be a DM candidate. We refer to it as the flavonic dark matter (FDM) \cite{Abbas:2023ion}. \\
The flavon attains VEV  as $\langle \chi  \rangle= f/ {\sqrt{2}}$, leading to $\mathcal{Z}_{\rm N}$ symmetry breaking.  
The axial flavon mass is expressed as \cite{Abbas:2023ion},
\begin{equation} 
\label{mphi1}
 m_a^2={1\over8} |\lambda| \tilde N^2 \epsilon^{\tilde N-4} f^2.
\end{equation}
To achieve the cold dark matter density correctly, the mass of the flavonic dark matter is determined as,
\begin{equation}
\label{mphi2}
  m_a =3.4\times 10^{-3} {\rm eV} \left( 10^{12} {\rm GeV} \over \varphi_0 \right)^4,
\end{equation}
where $\varphi_0$ represents the initial misalignment of the axial flavon field from its true vacuum during the inflationary phase. For details, see \cite{Abbas:2023ion}.
Taking $\varphi_0= a_0 f/\tilde N$, the required axial flavon mass $m_a$, and VEV $f$ turn out to be,
\begin{equation}
\label{flav_mass}
    m_a = 0.88 \times 10^{16} \left( \epsilon^{\tilde N-4} \tilde N^4 \frac{|\lambda|}{a_0^2} \right)^{2/5} {\rm eV},
\end{equation}
\begin{equation}
\label{eq:v_F}
    f =2.5\times 10^7 \left( \frac{ \tilde N^{6} }{ a_0^8|\lambda| \epsilon^{\tilde N-4} } \right)^{1/10} {\rm GeV}.
\end{equation}
We utilize a framework based on $\mathcal{Z}_{\rm 8} \times \mathcal{Z}_{\rm 22}$ flavour symmetry that furnishes a set-up  to impart dark-matter candidate with simultaneously addressing the flavour problem.  Unlike the two prototypes discussed in the section \ref{section2} , the $\mathcal{Z}_{\rm 8} \times \mathcal{Z}_{\rm 22}$ flavour symmetry prohibits the generation of mass of the top quark through tree-level SM Yukawa operator rather, they are generated by the non-renormalizable dimension-5 operator.
\begin{table}[tp]
 \small
\begin{tabular}{|c|c|c|c|c|c|c|c|c|c|c|c|c|c|c|}
  \hline
  Fields             &        $\mathcal{Z}_8$                    & $\mathcal{Z}_{22}$ & Fields             &        $\mathcal{Z}_8$                    & $\mathcal{Z}_{22}$   & Fields             &        $\mathcal{Z}_8$                    & $\mathcal{Z}_{22}$    & Fields             &        $\mathcal{Z}_8$                    & $\mathcal{Z}_{22}$     & Fields             &        $\mathcal{Z}_8$                    & $\mathcal{Z}_{22}$        \\
  \hline
  $u_{R}$                 &   $ \omega^2$  &$ \omega^2$        &$c_{R}$                 &   $ \omega^5$  & $ \omega^5$    &$t_{R}$                 &   $ \omega^6$  & $ \omega^6$       & $d_{R}$                 &   $ \omega^3$  &     $\omega^{3} $           & $s_{R}$                 &   $ \omega^4$  &     $\omega^4 $           \\
  $b_{R}$                 &   $ \omega^4$  &     $\omega^4 $     &   $\psi_{L,1}^q$                 &    $ \omega^2$  &    $\omega^{10} $      & $\psi_{L,2}^q$                 &  $ \omega$  &     $\omega^{9} $       &  $\psi_{L,3}^q$                 &    $\omega^{7} $  &      $\omega^{7} $ & $\psi_{L,1}^\ell$                 &   $ \omega^3$  &    $\omega^3 $          \\
     $\psi_{L,2}^\ell$                  &   $ \omega^2$  &    $\omega^2 $    &   $\psi_{L,3}^\ell$                 &   $ \omega^2$  &    $\omega^2 $     &  $e_R$                 &   $\omega^{4} $  &     $\omega^{12} $          & $\mu_R$                 &  $\omega^7 $   &     $\omega^{7} $      &  $\tau_R $                 &   $ \omega^7$  &     $\omega^{21} $              \\
            $ \nu_{e_R} $                 &     $\omega^2 $    &     $1 $         & $   \nu_{\mu_R}$                 &     $\omega^5 $    &     $\omega^{3} $          &  $  \nu_{\tau_R} $                 &     $\omega^6 $    &     $\omega^{4} $        &   $\chi$                        & $ \omega$  &       $ \omega$       & $H$              &   1        &     1                  \\          
  \hline
 \end{tabular}
 \caption{Assignment of charges to the SM and the flavon fields under the $\mathcal{Z}_8$ and $\mathcal{Z}_{22}$ symmetries, where $\omega$ is the 8th or 22nd root of unity.}
 \label{tab_z8z22}
\end{table}
Following the charge assignments to the SM  and flavon fields under these symmetries, as given in table \ref{tab_z8z22}, the mass matrices for the charged fermions can be written as,
\begin{equation}
M_u  = \dfrac{v}{\sqrt{2}}
\begin{pmatrix}
y_{11}^u  \epsilon^8 &  y_{12}^u \epsilon^{5}  & y_{13}^u \epsilon^{4}    \\
y_{21}^u \epsilon^7     & y_{22}^u \epsilon^4  &  y_{23}^u \epsilon^{3}  \\
y_{31}^u  \epsilon^{5}    &  y_{32}^u  \epsilon^2     &  y_{33}^u  \epsilon 
\end{pmatrix},
M_d   = \dfrac{v}{\sqrt{2}}
\begin{pmatrix}
y_{11}^d  \epsilon^7 &  y_{12}^d \epsilon^6 & y_{13}^d \epsilon^6   \\
y_{21}^d  \epsilon^6  & y_{22}^d \epsilon^5 &  y_{23}^d \epsilon^5  \\
 y_{31}^d \epsilon^4 &  y_{32}^d \epsilon^3   &  y_{33}^d \epsilon^3
\end{pmatrix}, 
M_\ell =  \dfrac{v}{\sqrt{2}}
\begin{pmatrix}
y_{11}^\ell  \epsilon^9 &  y_{12}^\ell \epsilon^4  & y_{13}^\ell \epsilon^4   \\
y_{21}^\ell  \epsilon^{10}  & y_{22}^\ell \epsilon^5  &  y_{23}^\ell \epsilon^3  \\
 y_{31}^\ell \epsilon^{8}   &  y_{32}^\ell \epsilon^5   &  y_{33}^\ell \epsilon^3
\end{pmatrix}.
\end{equation}
The approximate expressions for masses of the quarks and leptons at the leading-order are  \cite{Abbas:2023ion},
\begin{align}
\label{eqn:mass_z8z22}
\{m_t, m_c, m_u\} &\simeq \{|y_{33}^u| \epsilon , ~ \left |y_{22}^u  \right|  \epsilon^4 ,
~ \left |y_{11}^u \right| \epsilon^8\}v/\sqrt{2}  ,\nonumber \\ 
\{m_b, m_s, m_d\} & \simeq \{|y_{33}^d| \epsilon^3, ~ \left |y_{22}^d \right| \epsilon^5,
  \left |y_{11}^d \right| \epsilon^7\}v/\sqrt{2} ,\\ \nonumber 
\{m_\tau, m_\mu, m_e\} & \simeq \{|y_{33}^l| \epsilon^3, ~ \left|y_{22}^l\right| \epsilon^5, ~  \left |y_{11}^l \right| \epsilon^9\}v/\sqrt{2}. \nonumber 
\end{align}
The quark mixing angles are determined as  \cite{Abbas:2023ion},
\begin{eqnarray}
\sin \theta_{12}  \simeq& |V_{us}| \simeq \left|{y_{12}^d \over y_{22}^d}  -{y_{12}^u \over y_{22}^u}  \right| \epsilon, 
\sin \theta_{23}  \simeq& |V_{cb}| \simeq  \left|{y_{23}^d \over y_{33}^d}   -{y_{23}^u \over y_{33}^u}   \right| \epsilon^2, \nonumber \\
\sin \theta_{13}  \simeq& |V_{ub}| \simeq  \left|{y_{13}^d \over y_{33}^d}    -{y_{12}^u y_{23}^d \over y_{22}^u y_{33}^d}      
- {y_{13}^u \over y_{33}^u}   \right|   \epsilon^3.
\end{eqnarray}

For $\mathcal{Z}_{\rm 8} \times \mathcal{Z}_{\rm 22}$ flavour symmetry, the value of $\tilde N$ is determined to be 88, which using in equations \ref{flav_mass} and \ref{eq:v_F} result in,
\begin{align}
f  \approx 1.0 \times 10^{14}\, \rm{GeV}, ~~\text{and}~~
m_a \approx   1.9 \times 10^{-3} \,  \rm{eV},
\end{align}
assuming $\epsilon=0.225$ with $|\lambda|=1$ and $a_0=1$. To ensure the prolonged stability of flavonic dark matter, its decay into electrons and positrons pair  must be prohibited, implying $m_a < 2 m_e$. This condition sets a rigorous constraint on the parameter $\tilde N$ and VEV $f$ for the existence of flavonic dark matter, specifically requiring that \cite{Abbas:2023ion},
\begin{equation} \label{Nmin}
\tilde N > 53, ~~\mbox{and} ~~ f > 4 \times 10^{11} \mbox{GeV}.
\end{equation}
The decays of FDM to neutrino is highly suppressed due to coupling of the neutrinos to DM being of the order of $\epsilon^{20}$. The axial coupling of photon to DM turns out to be $g_{\varphi\gamma\gamma} = \frac{\alpha}{2 \pi v_F} \frac{5}{3}$, which results in the corresponding decay width $\Gamma (a \rightarrow \gamma \gamma) = 1.69 \times 10^{-68} \rm GeV$ \cite{Abbas:2023ion}. These results suggest that the lifetimes of these processes exceed the age of the universe, indicating the stability of flavonic dark matter.
\begin{figure}[h]
    \centering
\includegraphics[width= 0.40\linewidth]{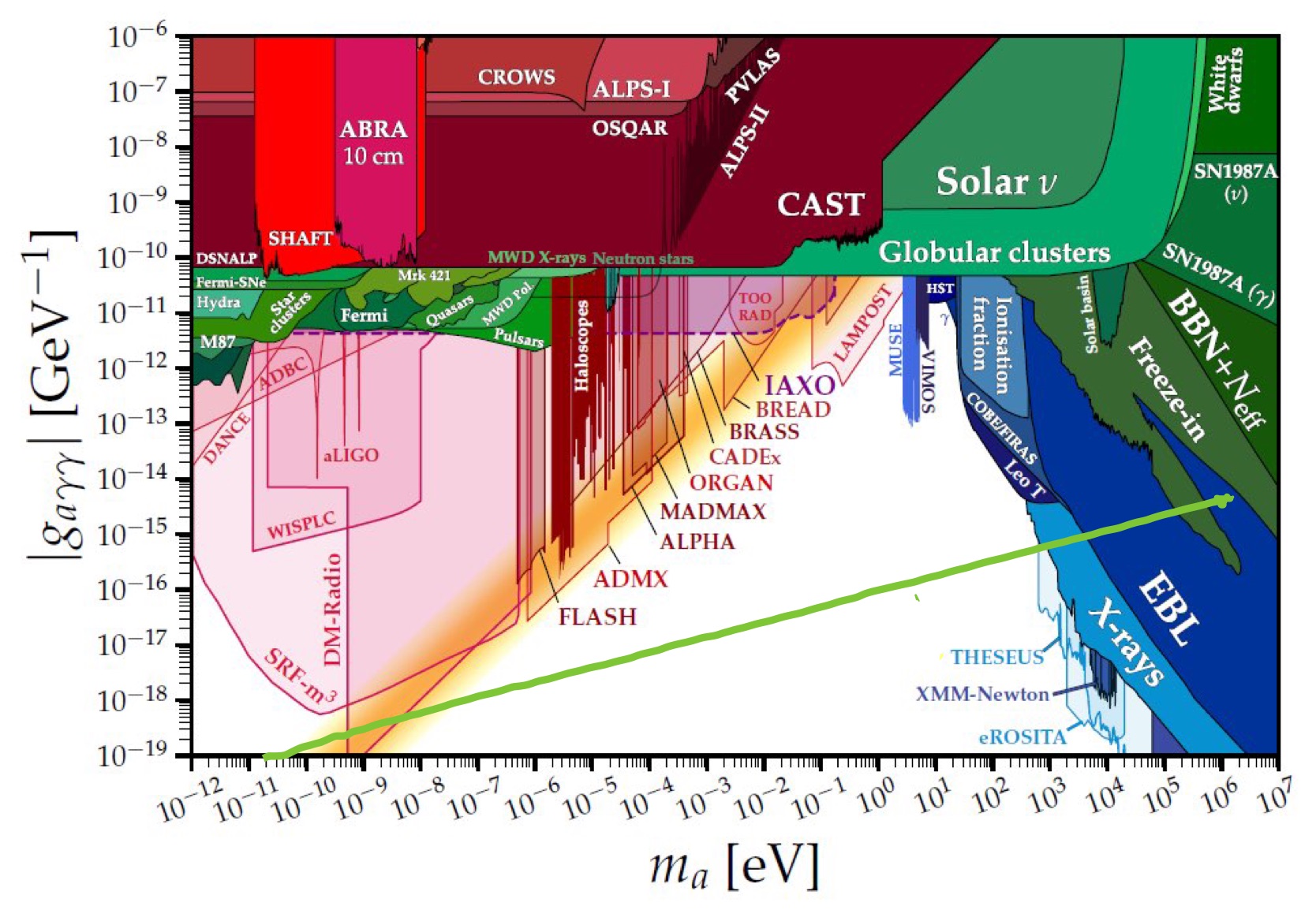} 
  \caption{ Predicted mass ranges for detection of flavonic dark matter (depicted by the thick green line) and axion-like particles. \cite{Antel:2023hkf}. }
    \label{fig:fips}
\end{figure}
The predicted photon coupling versus the flavonic dark matter (FDM) mass is presented in figure \ref{fig:fips}. It reveals that FDM masses exceeding approximately 1 keV, corresponding to $\tilde N < 67$ and $v_F < 4 \times 10^{12}$ GeV, are excluded. This exclusion aligns with recent constraints from INTEGRAL/SPI data \cite{INT}. Notably, these predictions also coincide with that of the GUT-scale QCD axion mass at around $10^{-9}$  eV, leading to a prospect for investigations in future \cite{DMRadio:2022jfv}. The mass range above the keV level can be explored further through the upcoming THESEUS experiment \cite{Thorpe-Morgan:2020rwc}. 
\section{The framework of the VEV hierarchy}
\label{sec4}
In this section, we turn up to an approach that stands in stark contrast to the FN mechanism relying on $\mathcal{Z}_{\rm N} \times \mathcal{Z}_{\rm M}$ flavour symmetry but shares the use of discrete symmetries. The Hierarchical VEVs model (HVM) firstly proposed in \cite{Abbas:2020frs} redefines the flavour problem by incorporating  three Abelian  discrete symmetries  $\mathcal{Z}_2$, $\mathcal{Z}_2^\prime$, $\mathcal{Z}_2^{\prime \prime} $ and six gauge singlet scalar fields $\chi_i, (i=1-6)$, whose VEVs account for both the fermionic mass spectrum and flavour mixing. The masses of the SM fermions are produced by the following dimension-5 operators,

\bea
\label{mass2}
{\mathcal{L}} &=& \dfrac{1}{\Lambda }\Bigl[  y_{ij}^u  \bar{\psi}_{L_i}^{q}  \tilde{\varphi} \psi_{R_i}^{u}   \chi _i +     
   y_{ij}^d  \bar{\psi}_{L_i}^{q}   \varphi \psi_{R_i}^{d}  \chi _{i+3}   +   y_{ij}^\ell  \bar{\psi}_{L_i}^{\ell}   \varphi \psi_{R_i}^{\ell}  \chi _{i+3} \Bigr]  
+  {\rm H.c.}.
\eea
The hierarchy in the masses and mixing patterns  is explained by  VEVs hierarchy in a way that  $ \langle \chi _4 \rangle \sim \langle \chi _1 \rangle $, $ \langle \chi _2 \rangle >> \langle \chi _5 \rangle $, $ \langle \chi _3 \rangle >> \langle \chi _6 \rangle $, $ \langle \chi _{3} \rangle >> \langle \chi _{2} \rangle >> \langle \chi _{1} \rangle $, and  $ \langle \chi _6 \rangle >> \langle \chi _5 \rangle >> \langle \chi _4 \rangle $. 

To account for the neutrino masses, an additional gauge singlet scalar field $\chi_7$ is added, and  neutrino mass patterns are recovered through the type-\rom{1} seesaw mechanism. The HVM framework, however, does not offer any explanation to the leptonic mixing angles. This problem is addressed by a standard HVM (SHVM)\cite{Abbas:2023dpf}, which provides remarkably precise predictions of leptonic mixing angles by linking them to the Cabibbo angle in the form of inequalities. 

The realization of SHVM involves enlarging the HVM framework  by extending the SM through the generic $\mathcal{Z}_{\rm N} \times \mathcal{Z}_{\rm M} \times \mathcal{Z}_{\rm P}$ symmetry. To produce the flavour structure of the SM  assuming normal hierarchy  for neutrino mass patterns, one must ensure that  $\rm N= 2$, $\rm M\geq 3$,  and $\rm P \geq 14$ \cite{Abbas:2023dpf}. We take the $\mathcal{Z}_2 \times \mathcal{Z}_4 \times \mathcal{Z}_{14} $ symmetry, and the charges to different SM fermionic and scalar fields under this symmetry are assigned, as shown in table \ref{tab:charges_shvm}.

\begin{table}[h]
\begin{center}
\begin{tabular}{|c|c|c|c||c|c|c|c||c|c|c|c||c|c|c|c|c|}
  \hline
  Fields                               &   $\mathcal{Z}_2$  &  $\mathcal{Z}_4$   &  $\mathcal{Z}_{14}$   & Fields   &  $\mathcal{Z}_2$   &  $\mathcal{Z}_4$ &  $\mathcal{Z}_{14}$ & Fields   & $\mathcal{Z}_2$  & $\mathcal{Z}_4$  &  $\mathcal{Z}_{14}$  & Fields  &  $\mathcal{Z}_2$   &  $\mathcal{Z}_4$  &  $\mathcal{Z}_{14}$ \\
  \hline
 $u_{R}$                        &     -   &     $ 1$          &    $\omega^{11}$ & $d_{R} $, $ s_{R}$, $b_{R}$    &     +    & $ 1$        &    $\omega^{12}$  & $ \psi_{L_3}^{q} $       &    +     &  $ 1$    &   $\omega^2$   &  $\tau_R$      &   +  &  $\omega^3$      &     $\omega$             \\
  $c_{R}$                       &     +   &    $ 1$          &    $\omega^6$   &  $\chi _4$                         &      +  &  $ 1$      &  $\omega^{13}$  &  $ \psi_{L_1}^\ell $                          &     +   & $\omega^3$     &  $\omega^{12}$                          &   $\nu_{e_R}$   &    +   &   $\omega$     &      $\omega^{8}$        \\
   $t_{R}$                        &     +   &    $1$         &    $\omega^{4}$   & $\chi _5$                         &      +  &  $1$    &  $\omega^{11}$  &  $ \psi_{L_2}^{\ell} $     &      +  & $\omega^3$      &  $\omega^{10}$     & $\nu_{\mu_R}$                   &     -  &   $\omega$    &   $\omega^3$                    \\
  $\chi _1$                        &      -  &   $ 1 $      &    $\omega^2$    &   $\chi _6$                          &      +  &  $ 1$      &   $ \omega^{10}$    &   $ \psi_{L_3}^{ \ell} $       &    +     &  $\omega^3$    &   $\omega^6$                                 & $\nu_{\tau_R}$                    &     -   &  $\omega$     &   $\omega^3$          \\
  $\chi _2$                   & +     &       $ 1$      &  $\omega^5$   & $ \psi_{L_1}^q $                          &      +  &  $1$      &  $\omega^{13}$  & $e_R$    &      -   &   $\omega^3$       &    $\omega^{10}$      &  $\chi_7 $                          &      -   &  $\omega^2$   &     $\omega^8$                                               \\
   $\chi _3$                  &    +   &       $ 1$    & $ \omega^2$        & $ \psi_{L_2}^{q} $     &      +  & $1 $      &  $\omega$  &   $ \mu_R$     &   +  & $\omega^3$       &     $\omega^{13}$      &  $ \varphi $                           &      +  &1     &   1                                            \\
  \hline
     \end{tabular}
\end{center}
\caption{Charges assigned to SM fermions  and  scalar fields under $\mathcal{Z}_2$, $\mathcal{Z}_4$  and $\mathcal{Z}_{14}$ symmetries, considering normal mass ordering for neutrinos.}
 \label{tab:charges_shvm}
\end{table} 

The mass patterns of charged fermions are obtained through dimension-5 operators in the Lagrangian given in equation \ref{mass2}, considering the VEVs pattern of  $\chi_i$ fields as discussed above. The corresponding mass matrices are,
\begin{align}
\label{mUD}
\M_\U & =   \dfrac{ v }{\sqrt{2}} 
\begin{pmatrix}
y_{11}^u  \epsilon_1 &  0  & y_{13}^u  \epsilon_2    \\
0    & y_{22}^u \epsilon_2  & y_{23}^u  \epsilon_5   \\
0   &  y_{32}^u  \epsilon_6    &  y_{33}^u  \epsilon_3 
\end{pmatrix},  
\M_\D = \dfrac{ v }{\sqrt{2}} 
 \begin{pmatrix}
  y_{11}^d \epsilon_4 &    y_{12}^d \epsilon_4 &  y_{13}^d \epsilon_4 \\
  y_{21}^d \epsilon_5 &     y_{22}^d \epsilon_5 &   y_{23}^d \epsilon_5\\
    y_{31}^d \epsilon_6 &     y_{32}^d \epsilon_6  &   y_{33}^d \epsilon_6\\
\end{pmatrix},
\M_\ell =\dfrac{ v }{\sqrt{2}} 
  \begin{pmatrix}
  y_{11}^\ell \epsilon_1 &    y_{12}^\ell \epsilon_4  &   y_{13}^\ell \epsilon_5 \\
 0 &    y_{22}^\ell \epsilon_5 &   y_{23}^\ell \epsilon_2\\
   0  &    0  &   y_{33}^\ell \epsilon_2 \\
\end{pmatrix},
\end{align} 
The masses of the charged fermions are retained as \cite{Abbas:2023dpf},
\begin{eqnarray}
\label{mass1a}
m_t  &=& \ \left|y^u_{33} \right| \epsilon_3 v/\sqrt{2}, ~
m_c  = \   \left|y^u_{22} \epsilon_2  - \dfrac{y_{23}^u  y_{32}^u \epsilon_5 \epsilon_6 }{y_{33}^u \epsilon_3} \right|  v /\sqrt{2} ,~
m_u  =  |y_{11}^u  |\,  \epsilon_1 v /\sqrt{2},\nonumber \\
m_b  &\approx& \ |y^d_{33}| \epsilon_6 v/\sqrt{2}, 
m_s  \approx \   \left|y^d_{22} - \dfrac{y_{23}^d  y_{32}^d}{y_{33}^d} \right| \epsilon_5 v /\sqrt{2} ,\nonumber \\
m_d  &\approx&  \left|y_{11}^d - {y_{12}^d y_{21}^d \over  y^d_{22} - \dfrac{y_{23}^d  y_{32}^d}{y_{33}^d} }   -
{{y_{13}^d (y_{31}^d y_{22}^d - y_{21}^d y_{32}^d )-y_{31}^d  y_{12}^d  y_{23}^d } \over 
{ (y^d_{22} - \dfrac{y_{23}^d  y_{32}^d}{y_{33}^d})  y^d_{33}}}   \right|\,  \epsilon_4 v /\sqrt{2},\nonumber \\
m_\tau  &\approx& \ |y^\ell_{33}| \epsilon_2 v/\sqrt{2}, ~
m_\mu  \approx \   |y^\ell_{22} | \epsilon_5 v /\sqrt{2} ,~
m_e  =  |y_{11}^\ell   |\,  \epsilon_1 v /\sqrt{2}.
\end {eqnarray}
The quark mixing angles are obtained as,
\begin{align}
\sin \theta_{12} &\simeq \left| \frac{y_{12}^d \epsilon_4}{y_{22}^d \epsilon_5} \right| = \frac{\epsilon_4}{\epsilon_5}, \quad
\sin \theta_{23} \simeq \left| \frac{y_{23}^d \epsilon_5}{y_{33}^d \epsilon_6} - \frac{y_{23}^u \epsilon_5}{y_{33}^u \epsilon_3} \right| \approx \left| \frac{y_{23}^d}{y_{33}^d} \right| \sin \theta_{12}^{2}, \\
\sin \theta_{13} &\simeq \left| \frac{y_{13}^d \epsilon_4}{y_{33}^d \epsilon_6} - \frac{y_{13}^u \epsilon_2}{y_{33}^u \epsilon_3} \right| \geq \left| \frac{y_{13}^d}{y_{33}^d} \right| \sin \theta_{12}^{3} - \left| \frac{y_{13}^u}{y_{33}^u} \right| \frac{m_c}{m_t} \approx \sin \theta_{12}^{3} - 2 \frac{m_c}{m_t},
\end{align}
where $\epsilon_i = \dfrac{\langle \chi _{i} \rangle }{\Lambda}$, $\epsilon_i<1$, and we made the assumptions that $\epsilon_4/\epsilon_5 = \sin \theta_{12} = 0.225$, $\epsilon_5/\epsilon_6 = \sin \theta_{23}$, $|y_{13}^d/y_{33}^d| \approx 1$, and $|y_{13}^u/y_{33}^u| \approx 2$. Consequently, the  SHVM predicts the quark mixing angle $\sin \theta_{13}$ to be approximately $\sin \theta_{12}^3 - 2(m_c/m_t)$, aligning well with its experimentally observed value.
\subsection{Neutrino masses and leptonic mixing parameters}
To recover neutrino masses, we add three right-handed neutrinos $\nu_{eR}$, $\nu_{\mu R}$, $\nu_{\tau R}$ to the SM and write the corresponding Yukawa Lagrangian as,
\begin{eqnarray}
\label{mass5}
-{\mathcal{L}}_{\rm Yukawa}^{\nu} &=&      y_{ij}^\nu \bar{ \psi}_{L_i}^\ell   \tilde{\varphi}  \nu_{f_R} \left[  \dfrac{ \chi_i \chi_j (\text{or}~ \chi_i  \chi_j^\dagger)}{\Lambda^2} \right] +  {\rm H.c.}. 
\end{eqnarray}
The Dirac mass matrix turn out to be,
\begin{equation}
\label{NM}
\M_{\D} = \dfrac{v}{\sqrt{2}}  
\begin{pmatrix}
y_{11}^\nu   \epsilon_1 \epsilon_7   &  y_{12}^\nu   \epsilon_4 \epsilon_7  & y_{13}^\nu  \epsilon_4  \epsilon_7 \\
0   & y_{22}^\nu  \epsilon_4  \epsilon_7 &  y_{23}^\nu  \epsilon_4  \epsilon_7 \\
0   &   y_{32}^\nu  \epsilon_5  \epsilon_7   &  y_{33}^\nu  \epsilon_5  \epsilon_7
\end{pmatrix},
\end{equation}
which results in following mass expressions for neutrinos,
\begin{eqnarray}
\label{mass1b}
m_3  &\approx&  |y^\nu_{33}|  \epsilon_5 \epsilon_7 v/\sqrt{2}, 
m_2  \approx     \left|y^\nu_{22} - \dfrac{y_{23}^\nu  y_{32}^\nu}{y_{33}^\nu} \right|  \epsilon_4 \epsilon_7 v /\sqrt{2},
m_1  \approx  |y_{11}^\nu  |\,  \epsilon_1 \epsilon_7 v /\sqrt{2}.
\end {eqnarray}
For the benchmark values of $y_{ij}^\nu$ obtained through our numerical analysis \cite{Abbas:2023dpf}, the masses of neutrinos turn out to have the values $\{m_3,m_2,m_1\} = \{5.05 \times 10^{-2}, 8.67 \times 10^{-3}, 2.67 \times 10^{-4} \}\, \text{eV}$.
The predictions for leptonic mixing are pivotal in SHVM. Assuming all the couplings to be of order one, the leptonic mixing angles can be expressed in terms of the Cabibbo angle in the following way \cite{Abbas:2023dpf},
\begin{eqnarray}
\sin \theta_{12}^\ell  &\simeq&  \left|-{y_{12}^\nu  \over y_{22}^\nu } +{y_{12}^\ell \epsilon_4  \over y_{22}^\ell  \epsilon_5}+  {y_{23}^{\ell *} y_{13}^\nu \epsilon_4  \over y_{33}^\ell y_{33}^\nu  \epsilon_5}   \right| \geq   \left|-{y_{12}^\nu  \over y_{22}^\nu }   \right| -   \left| {y_{12}^\ell   \over y_{22}^\ell  } +  {y_{23}^{\ell *} y_{13}^\nu   \over y_{33}^\ell y_{33}^\nu  }  \right|  {\epsilon_4  \over   \epsilon_5}\geq 1 - 2 \sin \theta_{12}, \\ \nonumber
\sin \theta_{23}^\ell  &\simeq&  \left|{y_{23}^\ell  \over y_{33}^\ell } - {y_{23}^\nu \epsilon_4  \over y_{33}^\nu  \epsilon_5} \right| \geq   \left|{y_{23}^\ell  \over y_{33}^\ell }   \right| -   \left| {y_{23}^\nu   \over y_{33}^\nu  }  \right|   { \epsilon_4  \over  \epsilon_5}\geq 1 -  \sin \theta_{12}, \\ \nonumber
\sin \theta_{13}^\ell    &\simeq& \left|    -{y_{13}^\nu \epsilon_4  \over y_{33}^\nu  \epsilon_5} + {y_{13}^\ell   \epsilon_5  \over y_{33}^\ell  \epsilon_2 }    \right| \geq  \left|    - {y_{13}^\nu   \over y_{33}^\nu  }  \right|  { \epsilon_4  \over   \epsilon_5}-  \left| {y_{13}^\ell     \over y_{33}^\ell   }  \right| { \epsilon_5  \over   \epsilon_2}  \geq \sin \theta_{12} - \frac{m_s}{m_c},
\end{eqnarray}
where $ m_s /m_c =      \epsilon_5  /   \epsilon_2    $.
The latest measurement of Cabibbo angle yields $\sin \theta_{12}    =  (0.225 \pm 0.00067)$ \cite{pdg22}. It is clear from above equations that $\sin \theta_{12}^\ell $  as well as $\sin \theta_{23}^\ell $ shares the same precision of the Cabibbo angle's experimental precision. Using the Cabibbo angle measurement, leptonic mixing angles are obtained as $\sin \theta_{12}^\ell  = 0.55 \pm 0.00134$ and $\sin \theta_{23}^\ell  = 0.775 \pm 0.00067$. 

Using running masses of strange and charm quarks at 1 TeV ($m_c = 0.532^{+0.074}_{-0.073}$ GeV, $m_s = 4.7^{+1.4}_{-1.3} \times 10^{-2}$ GeV) \cite{Xing:2007fb}, we obtain $\sin \theta_{13}^\ell = (0.1169 - 0.1509)$. However, the experimental lower end excludes $(0.1169 - 0.1413)$, and the theory rules out the upper end $(0.1509 - 0.1550)$. Thus, the final conclusive range for $\sin \theta_{13}^\ell$ is predicted to be $0.1413 - 0.1509$. The upcoming neutrino experiments like DUNE, Hyper-Kamiokande, and JUNO may test these predictions \cite{Huber:2022lpm}.\\
The SHVM rules out the possibility of neutrino masses originating from Majorana-type neutrinos, as the charge assignment of SHVM forbids both the Majorana Lagrangian and the Weinberg operators \cite{Abbas:2023dpf}. Consequently, the SHVM predicts neutrinos to be of the Dirac-type \cite{Abbas:2023dpf}.

\section{ The dark technicolour paradigm }
\label{sec5}
A UV completion of the SHVM can be obtained through the dark-technicolour framework  \cite{Abbas:2023bmm}.  The dark-technicolour model consists of the symmetry $\mathcal{S} =  SU(3)_c \times SU(2)_L \times U(1)_Y \times SU(\rm{N}_{\rm{TC}}) \times SU(\rm{N}_{\rm{DTC}}) \times SU(\rm{N}_{\rm{F}})  $ symmetry, where TC stands for technicolour, DTC is for the  dark-technicolour, and $\rm F$ is   a strong dynamics of vector-like fermions  \cite{Abbas:2020frs}.   The TC quarks transform under the symmetry    $\mathcal{S} $ as \cite{Abbas:2020frs},
\begin{eqnarray}
{T_L}^i  &\equiv&   \begin{pmatrix}
T_i  \\
B_i
\end{pmatrix}_L:(1,2,0, \rm{N}_{ TC},1,1),  
T_{i, R} : (1,1,1, \rm{N}_{ TC},1,1), \\ \nonumber
 B_{i, R}  &:& (1,1,-1, \rm{N}_{ TC},1,1), 
\end{eqnarray}
where electric charges $+\frac{1}{2}$ for $T$ and $-\frac{1}{2}$ for $B$. 

The $\rm DTC$ dynamics has the following fermions,
\begin{eqnarray}
 D  &\equiv& C_{i,  L,R}: (1,1, 1,1, \rm{N}_{\rm DTC},1),~ S_{i, L,R} : (1,1,-1,1, \rm{N}_{\rm DTC},1), 
\end{eqnarray} 
where   electric charges of the quark $\mathcal C$ is  $+\frac{1}{2}$, and that of the  $\mathcal S$ is  $-\frac{1}{2}$.   

The vector-like fermions of  $ SU(\rm N_{\rm F})$ dynamics transform as,
\begin{eqnarray}
F_{L,R} &\equiv &U_{L,R}^i \equiv  (3,1,\dfrac{4}{3},1,1, \rm{N}_{\rm F}),
D_{L,R}^{i} \equiv   (3,1,-\dfrac{2}{3},1,1,\rm{N}_{\rm F}),  \\ \nonumber 
N_{L,R}^i &\equiv&   (1,1,0,1,1,\rm{N}_{\rm F}), 
E_{L,R}^{i} \equiv   (1,1,-2,1,1,\rm{N}_{\rm F}).
\end{eqnarray}

There are three axial $U(1)_A^{\rm TC, DTC,  F}$ symmetries in the DTC paradigm which are  broken as  $ U(1)_A^{\rm TC, DTC, F} \rightarrow \mathcal{Z}_{2 \rm K_{\rm TC, DTC, F}}$ producing the generic   $\mathcal{Z}_{\rm N} \times \mathcal{Z}_{\rm M} \times \mathcal{Z}_{\rm P}$ flavour symmetry, where $\rm N= 2 \rm K_{\rm TC}$, $\rm M= 2 \rm K_{\rm DTC}$ and $\rm P= 2 \rm K_{\rm F}$  \cite{Harari:1981bs}.

\begin{figure}[h]
	\centering
 \includegraphics[width=\linewidth]{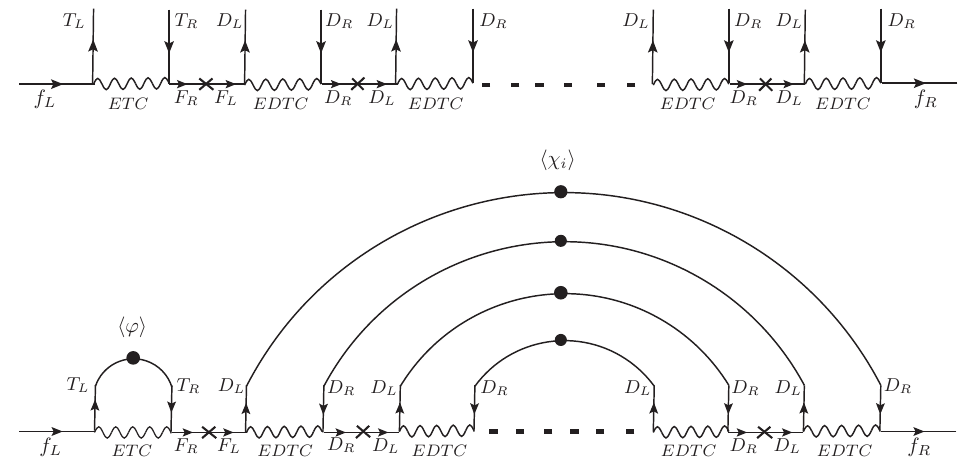}
    \caption{Feynman diagrams corresponding to charged fermion masses in the dark-technicolour paradigm. The diagram at the top depicts generic interactions of SM, TC, F, and DTC fermions, while diagram at the bottom shows the formation of fermionic condensates resulting in masses for charged SM fermions.}
 \label{fig1}	
 \end{figure}
 
The mass of a charged fermion is originated from the interactions given in fig. \ref{fig1} and  turns out to be \cite{Abbas:2023bmm},
\bea
\label{TC_masses}
m_{f} & = & y_f  \frac{\Lambda_{\text{TC}}^{3}}{\Lambda_{\text{ETC}}^2}  \dfrac{1}{\Lambda_{\text{F}}} \frac{\Lambda_{\text{DTC}}^{n_i + 1}}{\Lambda_{\text{EDTC}}^{n_i}} \exp(n_i k),~
\eea
where $n_i = 2,4, \cdots 2 n $ represents the number of fermions in a multi-fermion chiral condensate playing the role of the VEV $ \langle \chi_i \rangle$ in the SHVM \cite{Abbas:2020frs}.  The multi-fermion condensate can be parametrized as  \cite{Aoki:1983yy}, 
\be 
\label{VEV_h}
\langle  ( \bar{\psi}_L \psi_R )^n \rangle \sim \left(  \Lambda \exp(k \Delta \chi) \right)^{3n},
\ee
where $\Delta \chi$ shows the chirality of an operator, $k$ denotes a constant, and $\Lambda$ is  the scale of the  gauge theory.   From this, we conclude that,
\bea
\epsilon_i =  \dfrac{1}{\Lambda_{\text{F}}} \frac{\Lambda_{\text{DTC}}^{n_i + 1}}{\Lambda_{\text{EDTC}}^{n_i}} \exp(n_i k).
\eea

\begin{figure}[h]
	\centering
 \includegraphics[width=\linewidth]{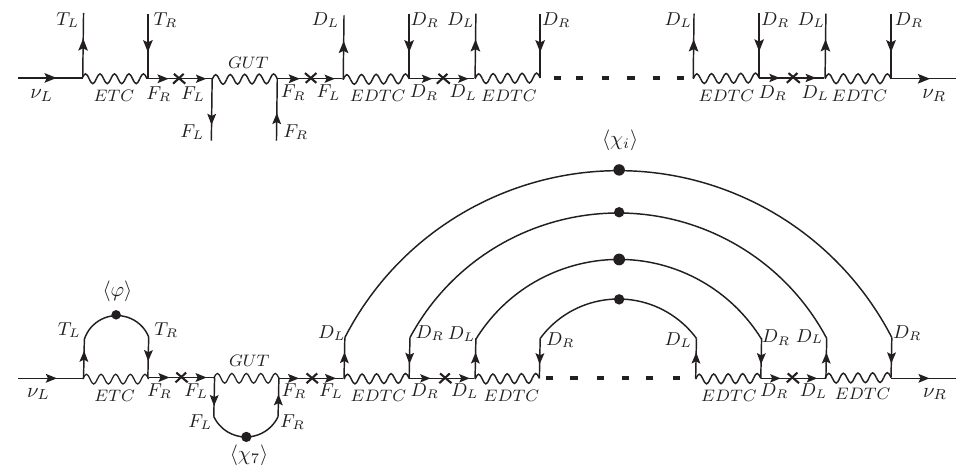}
    \caption{Feynman diagrams corresponding to the masses of neutrinos in the dark-technicolour paradigm. The diagram at the top shows generic interactions among the SM, TC, DTC gauge sectors, while the diagram at the bottom represents the formation of fermionic condensates leading to neutrino masses. }
 \label{fig2}	
 \end{figure}
 
 The neutrino masses are recovered by  the generic interactions given in figure \ref{fig2}, where it is assumed that  the TC and DTC sectors are accommodated in a GUT theory where the  interactions  between the $F_L$ and $F_R$ fermions are mediated by the gauge bosons of the GUT. The  VEV $\langle \chi_7 \rangle$ corresponds to the  chiral condensate $\langle \bar{F}_L F_R \rangle$.  The masses of neutrinos become,
\bea
\label{TC_nmasses}
m_{\nu} & = & y_f  \frac{\Lambda_{\text{TC}}^{3}}{\Lambda_{\text{ETC}}^2}  \dfrac{1}{\Lambda_{\text{F}}} \frac{\Lambda_{\text{DTC}}^{n_i + 1}}{\Lambda_{\text{EDTC}}^{n_i}} \exp(n_i k)  \dfrac{1}{\Lambda_{\text{F}}} \frac{\Lambda_{\text{F}}^{3}}{\Lambda_{\text{GUT}}^{2}} \exp(2 k).
\eea
For more details, see ref. \cite{Abbas:2023bmm}.

\section{ Summary }
\label{sec6}
In this work, we have discussed new approaches to address the flavour problem and dark matter together. The flavour bounds on the models based on  $\mathcal{Z}_{\rm N} \times \mathcal{Z}_{\rm M} $ flavour symmetries are extremely stringent in the phase-\rom{2} of the LHCb, and most of the parameter space is ruled out.

These new frameworks provides origins of the flavour problem and dark matter based  on discrete   $\mathcal{Z}_{\rm N} \times \mathcal{Z}_{\rm M} \times \mathcal{Z}_{\rm P}$ flavour symmetries.  Origin of these symmetries may lie in a technicolour paradigm.  We notice that these new frameworks result in a new class of dark matter referred  to as the  flavonic dark matter.

\section*{Acknowledgement}
This work is supported by the  Council of Science and Technology,  Govt. of Uttar Pradesh,  India through the  project ``   A new paradigm for flavour problem "  no. CST/D-1301,  and Science and Engineering Research Board,  Department of Science and Technology, Government of India through the project ``Higgs Physics within and beyond the Standard Model" no. CRG/2022/003237.  N. S. acknowledges the support through the INSPIRE fellowship by the Department of Science and Technology, Government of India.



\end{document}